\documentstyle[aps,epsfig]{revtex}

\begin{document}

\title{A microscopic model for mixed surfactant vesicles}

\author{ D. Duque $^1$,
P. Tarazona$^{1,2}$, and
E. Chac{\'o}n$^3$} 

\date{\today}

\maketitle

%\address{
$^1$Departamento de F{\'{\i}}sica Te\'orica de la Materia
Condensada,
Universidad Aut\'onoma de Madrid, E-28049 Madrid, Spain

%\address{
$^2$Instituto Nicol\'as Cabrera,
Universidad Aut\'onoma de Madrid, E-28049 Madrid, Spain

%\address{
$^3$Instituto de Ciencia de Materiales, Consejo Superior de
Investigaciones Cient{\'{\i}}ficas,  E-28049 Madrid, Spain

\begin{abstract}
A microscopic model which has proven useful in
describing amphiphilic aggregates as inhomogeneities
of a fluid is extended here to study the 
case of a two component surfactant mixture.
We have chosen an effective interaction between the
amphiphiles that mimics the mixture
of cationic--anionic surfactants.
In agreement with experiments, and other theoretical
approaches, we find regions where spherical vesicles
are stable, with a well defined radius.
The experimental dependence of the radius
on the chemical environment is also qualitatively
reflected in our model.
\end{abstract}

\section{Introduction}

 The structure and thermodynamical properties of fluid membranes composed
of amphiphilic molecules are of prime scientific and technological importance.
In particular the formation of vesicles has received very close attention 
due to their
pharmacological applications (microencapsulation) and their use as models 
for biological membranes. Usually, large vesicles are formed by 
intense mechanical shear (sonication), but they are a metastable phase 
which disappears after a time of hours or days. Only recently has been
found spontaneous formation
of long-lived vesicles, suggesting that the vesicles could be an
equilibrium state of the system
\cite{EL1,Kaler,Oberdiss1,Oberdiss2,GM31,GM32}.
We considerer here binary surfactant mixtures, which
are the main body of the experimental systems with this behavior
\cite{EL1,Kaler,Oberdiss1,Oberdiss2}, although there 
are also a few observations for systems with only one surfactant component
\cite{GM31,GM32}.
Note that the thermodynamic status of these systems is still
an open question. The 
spontaneous vesicles may be truly at equilibrium
or they may be a long-lived transient state to a lamellar phase 
\cite{Laughin}. It 
has also been argued that the existence of vesicles, and their size
distribution, may result from the edge energy and the dynamics of
growth \cite{Bettolo}.
Else, as an equilibrium property, 
that they arise from the entropy associated with 
the position and shape of the vesicles \cite{TERMO1}, rather than 
to the existence of a preferential curvature for an isolated vesicle.

We may distinguish two main kinds of two-component systems 
in which spontaneous formation of vesicles appear. 
In the first, the two amphiphiles have different self-assembly behavior.
One of them (a phospholipid molecule) is quasi insoluble in water and tends
to form 
lamellar phases, while the other (a surfactant molecule)
is soluble in water up to a critical surfactant concentration and then
tends to form micelles. 
One example of these systems is the mixture of octylglucoside and 
phosphatidylcholine 
\cite{EL1}. The second possibility is that both surfactants have similar 
self-assembly properties and its vesicular
behavior is due to the nature of the 
interaction
between the two species. The main examples of these systems are mixtures 
of two ionic surfactants, with oppositely charged head groups. The mixture
of the two types, in the solution, may form spontaneous vesicles, which
would not appear in pure solutions of either of the surfactants. 
Many experimental observations have been reported on these 
catanionic systems \cite{Kaler,Oberdiss1,Oberdiss2}.
In this work, we will only study this later kind of vesicular system, 
and search for the possible existence of spontaneous curvature,
and typical vesicle sizes, based on the thermodynamic stability
of single vesicles. We will neglect the effects of polydispersity and 
shape entropy. We will use the words {\it vesicle} and
{\it membrane} indistinctly 
to refer to bilayer aggregates; in each case, we will
make clear whether we are referring to a
closed vesicle or a planar membrane.

Recently 
Safran {\it et al} \cite{Safran1,Safran2} have treated this 
problem under the hypothesis that the existence of spontaneous
curvature for the vesicles is associated with the breaking of 
the composition symmetry between the two molecular layers in the 
membrane. In a semiempirical Landau theory, the free energy of a
spherical membrane of radius $R$ is 
written in terms of its curvature, ${\sc c} = 1/R$, and the 
relative concentrations of each species, $A$ and $B$, in each
molecular layer, $x_i^A=1-x_i^B$, with $i=1,2$, for the
two layers of the bilayer membrane. These relative concentrations
are described by an overall relative concentration
$$
\psi=\frac{1}{4}(x_1^A+x_2^A-x_1^B-x_2^B)=
\frac{1}{2}(x_1^A+x_2^A-1),
$$
which takes values $\pm 1$ for pure membranes (of $A$ or $B$ molecules)
and is zero for a membrane with symmetric concentrations of $A$ and $B$.
The second order parameter is a transverse relative concentration,
$$
\phi=\frac{1}{2}(x_2^A-x_1^A)=\frac{1}{2}(x_1^B-x_2^B),
$$
which measures the asymmetry between the two layers in the
membrane, being zero for symmetric membranes.
Since all the $x_i^\xi$ have to be between $0$ and $1$,
$\phi$ is constrained to be between $1/2-|\psi|$ and
$-1/2+|\psi|$.
For planar membranes, the free energy should be symmetric
in $\phi$, but there may be spontaneous symmetry breaking
when the membrane goes through a phase transition which
preferentially segregates the two species on different
sides of the bilayer membrane. For vesicles with
${\sc c} \neq 0$ there is a difference between the inner and the 
outer layer, which breaks the symmetry of the free energy
with respect to $\phi$.  The presence of a coupling term
proportional to ${\sc c} \phi$ favors the formation
of curved vesicles with asymmetric composition
$ {\sc c} \propto \phi $.
 The resulting phase diagram exhibits a line
of second-order transitions between the lamellar and vesicle phases,
with the highest critical temperature corresponding to the system with
a symmetric global composition $\psi=0$. 
At lower temperature, there is
a tricritical point where the transition
becomes first order. Although these semiempirical theories give a very
interesting and clarifying picture of these systems, they are parameter
dependent
and have a simplified functional dependence of the
free energy. To check the validity of the later and to relate
the parameters to molecular interactions requires microscopic
theories. Safran {\it et al} \cite{Safran1,Safran2} discussed
the use of their model to explain the
energetic stabilization of two-component vesicles,
for mixtures of amphiphiles which
differ only in their polar head groups. The curvature is
associated with the different concentrations of the two components
in the inner and outer monolayer of the membrane. Also,  
they assume that this different concentration between layers
is promoted by a interaction which gives phase separation
in the same layer. As a consequence, they choose a repulsive
interaction between the two different components.

We have recently developed a simple microscopic model
\cite{Somoza,Cilindro} for the study of aggregates of
monocomponent amphiphilic molecules in water, which
makes possible a
consistent thermodynamic treatment of the system by
taking account of the coexistence with the dilute
solution of amphiphiles.
The model considers only the positional and orientational 
distribution of the amphiphiles, 
ignoring internal degrees of freedom. The effective 
interactions between the amphiphiles
is described by a potential which includes the interaction with the
water (so that water molecules need not be explicitly taken into account).
This minimal model gives a qualitatively correct global phase diagram for
aggregates with planar, spherical, and cylindrical symmetry. The elastic
constants obtained are in good agreement with experimental results.
Depending of the interaction potential, the model exhibits coexistence
between a solution of free molecules,
micelles, and planar membranes.
This monocomponent model does not give spontaneous curvature for any
interaction potential and therefore the spherical vesicles are always 
unstable when compared with the micelles or with the planar membranes. 

The goal of the present 
work is to extend the above model to the study of stability and phase
behavior of two--component vesicles. As in the study of the elastic constants
\cite{Cilindro}, the self consistent thermodynamic
treatment, i.e. the thermodynamic
equilibrium between the solution
of amphiphiles and the vesicles, shows interesting
restrictions on the possible
values of the parameters used in semiempirical Landau 
theories.  In order to exclude other possible sources of
spontaneous curvature, we use a model in which the interactions are fully
symmetric. Thus, the pure component membrane made with $A$ molecules
is exactly equivalent to the pure $B$ membrane. In experimental
systems with catanionic surfactants this symmetry is not perfect, i.e.
the properties of the two types of pure component systems are not 
equal, but still they are much more similar than the properties of the
surfactant mixture.

\section{Density functional model}

In this section we extend our  previous approach
\cite{Somoza} to
the case of two different components, which we 
label $\xi = A$, $B$. 
The model is based on the description of the microscopic
or mesoscopic molecular aggregates in water,
as inhomogeneous density distributions, treated within the
density functional formalism.  We use an approximation
to the exact Helmholtz free energy density functional,
$ F[ \rho^{\xi} ]$, applied to a very simple molecular interaction model.
As the amphiphilic molecules are much larger that the water molecules,
our previous approach described
the interaction between them by a potential which represents 
an effective, water-mediated interaction ( and so, water molecules
need not be explicitly taken into account in the model).
The head part of an amphiphilic
molecule attracts water much more strongly that the lipid tail. 
This fact produces
an effective repulsion between the head of a molecule and
any other molecule (regardless of the orientation of the latter
because it displaces the water).
We consider the amphiphiles as uniaxial 
molecules without internal
degrees of freedom, other than the orientation of the head-tail 
direction, which is described 
by a unit vector ${\bf \hat{u}}$ along the molecular axis.
Let us consider a $\xi$--molecule at ${\bf r}_1$, with
orientation ${\bf \hat{u}_1}$, and a $\xi'$--molecule at ${\bf r}_2$ with
orientation ${\bf \hat{u}_2}$.
The effective pairwise potentials between them, with
the proper features, may be
written as
$ \Phi^{ \xi, \xi^{\small{'}} }
({\bf r}_2-{\bf r}_1,\hat{\bf u}_1,\hat{\bf u}_2)  =
\Phi_{0}^{ \xi, \xi^{\small{'}} }(r_{21}) +
\Phi_a^{ \xi, \xi^{\small{'}} }({\bf r}_{21},\hat{\bf u}_1,\hat{\bf u}_2)$.
The isotropic part $ \Phi_{0} $ only depends on the distance
between the molecular centers, $r_{21}=|{\bf r}_{21}|=
|{\bf r}_2-{\bf r}_1|$, and the anisotropic part, $ \Phi_{a} $,
may be written as a general expansion in spherical harmonics.
As we have argued in our previous papers on monocomponent
systems \cite{Somoza,Cilindro},
we need only include the terms
which do not couple the orientation of the two molecules.
The general form is
\begin{eqnarray}
\Phi_a^{ \xi, \xi^{\small{'}} }({\bf r}_{21},\hat{\bf u}_1,\hat{\bf u}_2) & = &
\sum_{i=1}^{\infty}
\Phi_{i}^{ \xi, \xi^{\small{'}} }(r_{21}) \times  \nonumber   \\  & &
 \left[ P_{i}(\hat{\bf u}_1 \cdot \hat{\bf r}_{21}) +
 P_{i}(-\hat{\bf u}_2 \cdot \hat{\bf r}_{21}) \right],
\label{Phi}
\end{eqnarray}
where ${\bf \hat{r}}_{21}= {\bf r}_{21}/r_{21}$ and
$P_i(x)$ are the Legendre polynomials.
This restriction of the model interactions has the
advantage that minimization with respect to 
orientational degrees of freedom can be done analytically. 

The Helmholtz free energy density functional
may be approximated by
\begin{eqnarray}
F[\rho^{A}({\bf r},\hat{\bf u}),\rho^{B}({\bf r},\hat{\bf u})]  
= F_{0}[\rho^{A}({\bf r}),\rho^{B}({\bf r})]  \nonumber \\
+ k_B T \sum_{ \xi = A,B} \int d{\bf r}_1 d\hat{\bf u}_1
\rho^{\xi}({\bf r}_1)  \alpha^{\xi}({
\bf r}_1,\hat{\bf u}_1)
\log(4  \pi \alpha^{\xi}({\bf r}_1,\hat{\bf u}_1))    \nonumber \\
+ {1 \over 2} \sum_{ \xi = A,B} \sum_{ \xi^{\small{'}} = A ,B}
\int d{\bf r}_1 d\hat{\bf u}_1
d{\bf r}_2  d\hat{\bf u}_2
\rho^{\xi}({\bf r}_1,\hat{\bf u}_1) 
\rho^{\xi^{\small{'}}}({\bf r}_2,\hat{\bf u}_2)
\Phi_a^{{\xi} \xi^{\small{'}} } ( {\bf r}_{21}, \hat{\bf u
}_1 , \hat{\bf u}_2 ).
\label{FF1}
\end{eqnarray}
The first term is the density functional for the isotropic 
reference fluid; it depends on the density distributions, 
$\rho^{\xi}({\bf r})$.
The second term is the rotational entropy and the third one is the mean
field contribution of the anisotropic interactions; both
depend on the position and molecular orientation distributions,
$\rho^{\xi}({\bf r},\hat{\bf u})$, with
$\alpha^{\xi}({\bf r},\hat{\bf u}) \equiv
\rho^{\xi}({\bf r},\hat{\bf u}) / \rho^{\xi}({\bf r})$.

As stated in the introduction we are interested in the 
study of a mixture of surfactants with similar 
self-assembly properties, so that the formation of vesicles is 
due to the interaction between the two species.
Within this scheme, it is logical to assume that
the $A$--$A$ interaction is equal to the $B$--$B$ one,
but different from the $A$--$B$ interaction:  
$ \Phi_a^{A A} = \Phi_a^{B B} \neq \Phi_a^{A B} $. 
We have fixed the shape of the potentials following the experience
gained in the monocomponent case. We consider that
the isotropic potential energy $\Phi_{0}^{ \xi \xi^{\small{'}}}(r)$
is essentially repulsive, since an attractive
contribution would produce the unrealistic
segregation of water-rich and amphiphilic-rich bulk fluid phases.
As the simplest choice, we use a repulsive isotropic interaction consisting
of a hard sphere (HS) potential,
$\Phi_{\mathrm{hs}}^{ \xi \xi^{\small{'}} }(r)$. By symmetry, the
sphere diameter, $d_{\mathrm{hs}}$, is taken to be identical
for the two species. 
We choose this hard sphere diameter as the unit length
and $k_B T = \beta^{-1}$ as the unit of energy. For the
anisotropic potential we have considered a single form,
irrespective of the species involved, changing only their prefactors.
The first coefficient function of the anisotropic potential given by 
eq 1 was taken to be an empty-core Yukawa potential,
$$
\Phi_{1}^{ \xi \xi^{\small{'}} }(r)=
{K^{ \xi \xi^{\small{'}} } \over r} exp(- \lambda \ (r-d_{\mathrm{hs}})), 
$$
for $r \geq d_{\mathrm{hs}}$, and null inside the hard core, $r < d_{\mathrm{hs}}$.
The next coefficient function in
eq 1 is taken to be proportional to the first,
$\Phi_2^{ \xi \xi^{\small{'}} }(r)= q \Phi_1^{ \xi 
\xi^{\small{'}} }(r)$, and all the rest 
coefficient functions are neglected.
Based in our previous work, we will consistently use 
$ \lambda d_{\mathrm{hs}} =2 $ and $ q = 0.5 $. The latter choice is made 
so that the monocomponent planar membrane will be stable
with respect to the micelles, in the
range of couplings we consider. In contrast,  
when $ q \leq 0.3 $ the micellar structures are more stable. 
Thus, the model only has two unprefixed adimensional parameters: 
$\tilde{K}= \beta K^{AA}/d_{\mathrm{hs}} = \beta K^{BB}/ d_{\mathrm{hs}} $, the 
measure of the amphiphilic character of the pure surfactants, 
and the ratio of like/dislike energies, given by $\epsilon=K_{AB}/K_{AA}$.
Note that within this model we could study the first kind of
vesicle systems, in which the two surfactant have different self-assembly
behavior, considering, e.g., $q =0$ for the surfactant component
and $q = 0.5$ for the phospholipid molecules. This
study is left for a later work.

For a given temperature, $T$, and
chemical potentials, $\mu^{A}, \mu^{B}$, we have to search for local minima of the
grand potential energy, with respect to
the distribution functions $\rho^{\xi}({\bf r}, \hat{\bf u})$.
Due to our simplified choice of anisotropic interactions,
the minimization with respect to $\alpha^{\xi}({\bf r},\hat{\bf u})$ may be
carried out analytically. Thus,
$$
\alpha^{\xi}({\bf r},{\bf \hat{u}})=\frac{1}{z^{\xi}}
                  \exp\left(
                      {\sum_{\xi^{\small{'}} =A,B }{\bf a}^{ \xi 
                  \xi^{\small{'}} }{\bf \hat{u}}+
                       \sum_{\xi^{\small{'}} =A,B} {\bf \hat{u}}^T 
                      \cal{B}^{ \xi \xi^{\small{'}} } {\bf \hat{u}} } \right),
$$
where the constants $z^{\xi}$ ensure the normalization condition of $\alpha^{\xi}$, i.e.
$\int \alpha^{\xi}({\bf r},{\bf \hat{u}}) d{\bf \hat{u}}=1$.
For each possible value of $\xi$ and $\xi^{\small{'}}$,
 ${\bf a}^{ \xi \xi^{\small{'}} }$ is a vector
arising from the first coefficient function of the
anisotropic potential and $\cal{B}^{\xi \xi'}$ is a tensor
associated to the second coefficient. Their values are given by:
$$
{\bf a}^{ \xi \xi^{\small{'}} }({\bf r}_1)=
    - \beta \int d{\bf r}_2 \rho^{\xi^{\small{'}}}({\bf r}_2) \Phi^{  
\xi \xi^{\small{'}} }_1(|{\bf r}_{1 2}|) {\bf \hat{r}}_{1 2}
$$
and 
$$
 {\cal B}^{  \xi \xi^{\small{'}} }({\bf r}_1)=
    - \beta \int d{\bf r}_2 \rho^{\xi^{\small{'}}}({\bf r}_2) \Phi^{  
    \xi \xi^{\small{'}} }_2(|{\bf r}_{1 2}|)
     \frac{3 {\bf \hat{r}}_{1 2}\otimes
    {\bf \hat{r}}_{1 2} - {\bf \bar{I}} }{2},
$$
where ${\bf \bar{I}}$ is the identity matrix and $\otimes$ is a direct
product. For planar and spherical symmetries the expressions
simplify and we only have to consider the modulus of
${\bf a}^{  \xi \xi^{\small{'}} }({\bf r})$ and the
largest eigenvalue of ${\cal B}^{  \xi \xi^{\small{'}} }({\bf r})$; these we will
call $a^{  \xi \xi^{\small{'}} }({\bf r})$ and $b^{  \xi 
\xi^{\small{'}} }({\bf r})$, respectively.

By substitution of the above into eq 2 we get a free energy density
functional, which is already minimized with respect to the
molecular orientations:
\begin{eqnarray}
F[\rho^{A}({\bf r}), \rho^{B}({\bf r})] &=&
\min\{ F[\rho^{A}({\bf r},\hat{\bf u}),
        \rho^{A}({\bf r},\hat{\bf u})] \ \}_{\alpha} = \nonumber \\
&=& \Delta F_{\mathrm{hs}}[\rho({\bf r})] +  
 k_B T \sum_{ \xi =A,B} \int d{\bf r}_1 \rho^{\xi}({\bf r}_1) 
\log(\rho^{\xi}({\bf r}_1)-1)     \label{FF2}  \\
&-& \sum_{ \xi =A,B} \int d{\bf r} \rho^{\xi}({\bf r}) 
\log \left(Q^{\xi}({\bf r})\right),
\nonumber 
\end{eqnarray}
where
$$
Q^{\xi}({\bf r})=\frac{1}{2}\int_{-1}^1 dx \exp
           \left(
              \sum_{\xi^{\small{'}} =A,B} \left\{ a^{ \xi 
\xi^{\small{'}} }({\bf r}) P_1(x)+
b^{ \xi \xi^{\small{'}} }({\bf r}) P_2(x) \right\}   \right),
$$
and $ \rho({\bf r}) = \rho^{A}({\bf r}) + \rho^{B}({\bf r}) $ is
the total density. In our calculations
the interaction part $ \Delta F_{\mathrm{hs}}[\rho({\bf r})] $ of the hard sphere
free energy is approximated by a well 
tested non-local density functional \cite{nonlocalhs}.

In absence of
external potentials, this density functional
always has a stationary grand-potential energy for a
homogeneous density distribution
$\rho^{\xi}({\bf r})=\rho_0^{\xi}$, for which   
${\bf a}^{ \xi \xi^{\small{'}} }({\bf r})$ and $ {\cal B}^{ \xi 
\xi^{\small{'}} } ({\bf r}) $ vanish. This state
represents a mixture of surfactant molecules with random orientations,
which in our mean field description is equivalent to a hard
spheres mixture. The densities of
this homogeneous solution are used to control the chemical
potentials, $\mu^{\xi}$, in our search
for inhomogeneous  density distributions. In this case the expression for
the chemical potentials is
$$
\beta\mu^{\xi}=\log(\rho_0^\xi)+\beta\mu_{\mathrm{hs}}(\rho_0),
$$
where $\mu_{\mathrm{hs}}(\rho_0)$ is the excess (over the ideal)
chemical potential of a hard sphere fluid at a total density
$\rho_0$.

  For low values of $\mu^{\xi}$ (and $\rho_0^{\xi}$), the homogeneous phase
represents a dilute solution of amphiphilic molecules in water and
we may find other local minima of the grand potential energy,
\begin{equation}
\Omega[\rho^{A}({\bf r}), \rho^{B}({\bf r})] = F[\rho^{A}({\bf r}), \rho^{B}({\bf r})] \ 
- \mu^{A} \ \int d{\bf r} \ \rho^{A}({\bf r})
- \mu^{B} \ \int d{\bf r} \ \rho^{B}({\bf r}),
\label{Omega}
\end{equation}
for non-uniform density distributions, which represent
the different types of molecular aggregates.

\section{ Results} 

We are interested in the vesicle-planar membrane phase diagram.
In this phase diagram we must 
require that the aggregates are in chemical equilibrium
with the dilute surrounding solution of free molecules (which we call
``{\it the bulk}''), i.e. that they be local minima of eq 4. 
Beyond of this condition of local stability,  we have to seek a
global stability, which means that the excess grand-potential
energy of the aggregate, with respect to the bulk solution, 
has to be exactly zero. This is a strong requirement, which comes
from the consistent thermodynamic treatment of the model, and it
sets important restrictions on the possible structures of the
aggregates.

Although the minimization of eq 4 can be carried out exactly, 
we have calculated the
coexistence line within a parameterized family of variational density profiles,
which improves the computational efficiency without important changes in the
results \cite{Somoza}. 
With this approach, it is easy to impose a well--defined
mean radius to the structure, whose
inverse is the curvature, $ {\sc c} $.
In order to search for vesicles, we impose spherical symmetry and minimize the
free energy of our system with respect to the density profiles.
As vesicles are characterized by two peaks 
corresponding to the two monolayers, a
useful parameterization of each of the two density profiles
is achieved by two Gaussians superimposed on an uniform background;
hence we will need four of them. The minimization
is made with respect to the twelve
parameters that characterize the Gaussians:
widths $\alpha_i^{\xi}$ (four parameters),
surface densities of the monolayers $\eta_i$ (two parameters),
relative compositions $x_i^\xi$ (two parameters),
radius $R_v^\xi$
(two parameters), and distances between layers  
$d^\xi$ (two parameters). In practice, the radius and the distance between
layers are almost always equal for both species.
Thus, our density distributions are given by:
\begin{eqnarray}
\rho^{\xi} (R) & = & \rho_{0}^{\xi} +
\eta_1 x_1^\xi \left(\frac{\alpha_1^\xi}{\pi}\right)^{1/2}
   \exp [-\alpha_1^\xi (R-(R_v^\xi-d^\xi))^2]
\nonumber \\  &+&
\eta_2 x_2^\xi \left(\frac{\alpha_2^\xi}{\pi}\right)^{1/2}
   \exp [-\alpha_2^\xi (R-(R_v^\xi+d^\xi))^2]
\label{param}
\end{eqnarray}
Here, the $x_i^\xi$ have the same meaning as in ref 10 and 11,
and from them we get directly the variables $\phi$ and
$\psi$ defined above.

In Figure 1 we show a typical profile obtained with our model
for a spherical vesicle. As can be seen, the profile exhibits
two layers of amphiphiles with different
composition. In this case the concentration of the 
minority component $B$ is higher
in the outer layer than the inner one and the majority component
has an opposite behavior. The orientational profiles are very similar to
those obtained in the monocomponent case, being saturated and
of opposite sign at the monolayers, thus indicating the different orientations
of the molecules.

\begin{figure}
\begin{center}
\epsfig{file=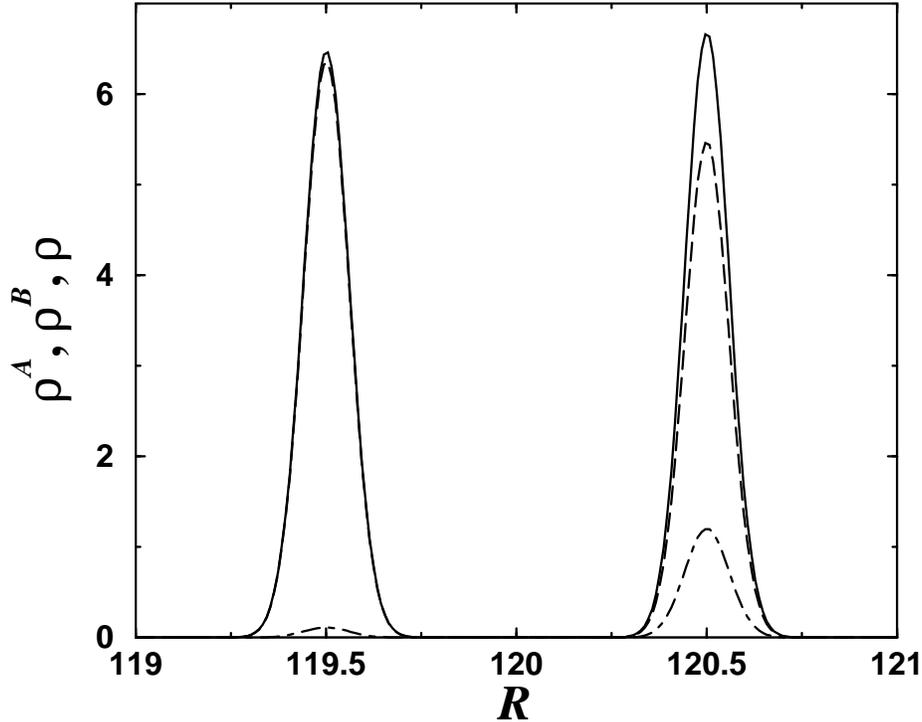,width=12cm}
\caption[]{ 
Density profiles
for a  spherical vesicle at equilibrium,
with bulk densities
$\rho_0^{A}=9\ 10^{-7}$, and $\rho_0^{B}=3.4\ 10^{-14}$.
Dashed line: density of the majority component $\rho^A(R)$.
Dot--dashed  line:
density of the minority component $\rho^B(R)$.
Solid line:
full density profile $\rho(R)$.
The interaction potential parameters are
$\tilde{K}=10$ and $\epsilon=1.4$.
The hard sphere diameter is used to provide the units of
distance and density.
}
\end{center}
\end{figure}

As first step we study the phase diagram 
for planar membranes, taking the
vesicle radius, $ R_{v} $, to infinity.
In Figure 2 we present the line of stability for
$\epsilon=1.4$, with $\tilde{K}$
being a function of the total density of surfactant in the bulk solution,
$ \rho_0= \rho^{A}_{0}+\rho^{B}_{0}$, 
for different values of the difference in the chemical potential
of the two species, $\beta \Delta \mu=\log(\rho^{A}_{0}/\rho^{B}_{0})$.
In the region below each line, the membranes have $\Delta\Omega > 0$, i.e. 
they are unstable and would  dissolve in the water. In the upper region
$ \Delta\Omega < 0 $, so that the planar membranes grow taking particles
from the bath; 
then the 
bulk density is depleted to the coexistence value. The main
qualitative change, with respect to the pure surfactant membranes
comes from the possibility that membranes have a spontaneous
symmetry breaking, which gives them asymmetric composition
$\phi \neq 0$. This effect happens as a second-order phase transition, 
at positions on the coexistence curves (the
places marked by the arrows in Figure 2)
which help in stabilizing the membranes and lower down the lines
in the figure. 
It is important to remark that this segregation is only possible when
there is a reasonable amount of the two species, $A$ and $B$, in 
the membrane. But the effect is obtained even if the bulk solution 
has strong majority of one of the species. The bilayer
membrane is able to concentrate the minority species by
many orders of magnitude of its bulk concentration, and then
it may segregate preferentially in one of the the two 
molecular layers.
Obviously, once the planar membrane undergoes the phase transition,
there is no symmetry reason to avoid spontaneous curvature of the
membrane into a closed vesicle.
This is the coupling, between the composition asymmetry and spontaneous curvature,
which acts to stabilize the vesicles proposed by
Safran {\sl et al} \cite{Safran1,Safran2}, and we may now analyze it
within our microscopic model.
The main difference here is that in our model this coupling is
fixed by the molecular interactions and by the requirements of
global thermodynamic stability, instead of being a free parameter
(assumed constant) in the effective Landau free energy.

\begin{figure}
\begin{center}
\epsfig{file=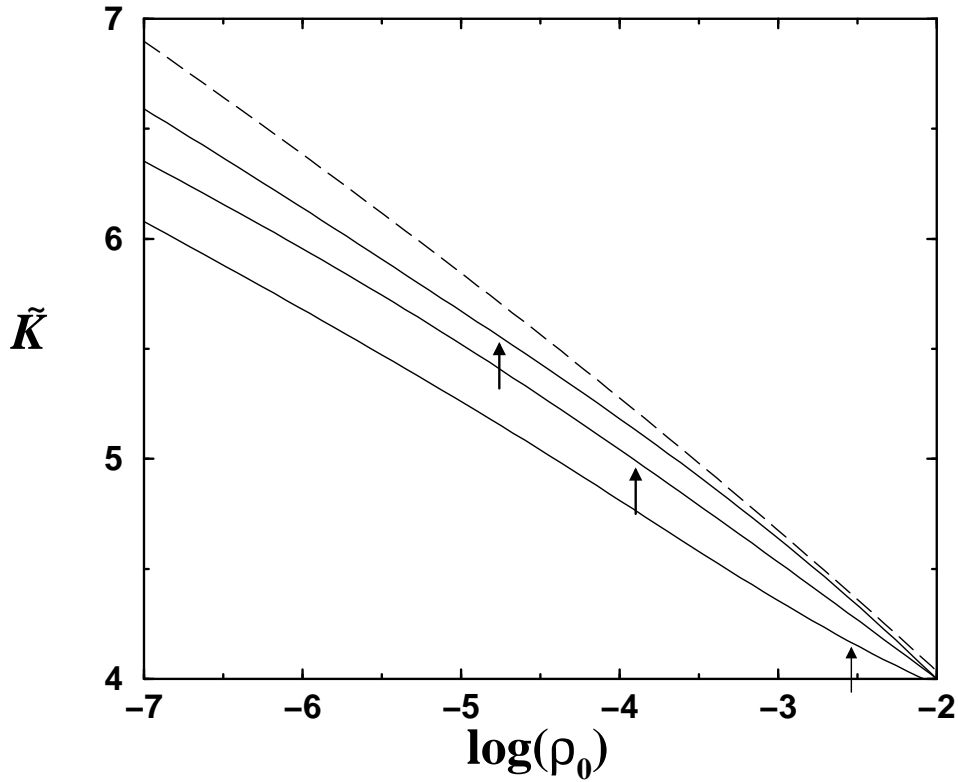,width=12cm}
\caption[]{
Equilibrium lines in a $\log(\rho_0)$ -- $\tilde{K}$ phase diagram
for planar membranes, with $\epsilon = 1.4$.
The dashed line gives the monocomponent limit
$ \Delta \mu \rightarrow \infty $.
The full lines show the coexistence line for several values
of $ \Delta \mu =  4, 6, 8 $, from bottom to top.
The arrows mark the points where
$\phi$ becomes different from zero.
}
\end{center}
\end{figure}
 
The first result of the consistency within our
microscopic model, in which we have assumed perfect symmetry between the two
types of surfactant, is that the spontaneous curvature  should
vanish not only for symmetric membranes, with $\phi=0$, but also
for membranes with preferential segregation of each surfactant
in one of the layers, $\phi \neq 0 $, but with equal global
concentration of the two species, $\psi=0$. Thus, in the Landau
free energy the coupling term has to go like
${\sc c} \phi f ( \psi ) $ (where f is an odd  function of $\psi$ that
vanishes at $\psi=0$), instead of being a constant times ${\sc c} \phi$,
as taken in ref 11. In real catanionic mixtures, this 
symmetry between the two components is not perfect, but still they
are probably closer to our ideally symmetric model than to the
the model with constant $f(\psi)$. 

The simplest Landau free energy we may write, with this requirement,
is  
$$
f=T_1 \phi^2 +T_2 \psi^2 + a_1 \phi^4 + a_2 \psi^4 +a_{1 2} \phi^2\psi^2+
\kappa[{\sc c}^2-2\gamma {\sc c} \phi \psi ]-2\psi\Delta\mu,
$$

\noindent where we have approximated the function $ f $ by
a linear function of $ \psi $.
The coefficients $a_1$, $a_2$, $a_{1 2}$,
$\kappa$ and $\gamma$ are
assumed to be positive constants. $T_1$ and $T_2$ correspond
to the reduced critical temperatures, which are negative below
critical points for the order parameters
$\phi$ and $\psi$ in the planar membranes (${\sc c} =0$). In the
effective free energy used by
MacKintosh and Safran \cite{Safran2}, the two temperatures 
$T_1$ and $T_2$ are taken to be identical, which would 
correspond to independent phase transitions in each monolayer,
with the preferential segregation of each type of surfactants
on uncorrelated patches in the two sides of the bilayer.
In our case (with the term $\gamma {\sc c} \phi \psi$ instead of
$\gamma {\sc c} \phi $) we require $T_1<T_2$, so that the
parameter $\phi$ goes through a phase transition when $T_1$
changes sign while $T_2$ is always positive.
In the case of catanionic surfactants our assumption is apparently
easier to justify, given
long ranged attraction between molecules of different types,
than the hypothesis of MacKintosh and Safran, which requires  
an effective attraction between molecules of the same species
in order to have segregation at each monolayer.
 
The minimization with respect to the curvature leads to 
an effective free energy

\begin{equation}
f=T_1 \phi^2 +T_2 \psi^2 + a_1 \phi^4 + a_2 \psi^4 +
(a_{1 2}+3 \kappa\gamma^2) \phi^2\psi^2 -2(\Delta\mu)\psi. 
\label{landau}
\end{equation}
There is a line of critical temperatures $T_c(\Delta \mu)$, starting
at $T_c(0)=T_1$,
below  which both ${\sc c}$ and $\phi$ become nonzero. 
In Figure 3, we present a sketch of equilibrium curvature
of vesicles, at fixed $T$, as a function of the
$\Delta \mu$ in
the free energy given by eq 6
---dot--dashed line---
and by the free energy
analyzed by MacKintosh and Safran \cite{Safran2}
---dashed line---.
The qualitative difference is the behavior at $\Delta \mu=0$
associated to the symmetry of our model.
In both cases, the general prediction of the Landau
theory is that there may exist a tricritical point at which
the transition becomes first order, through the
coupling with $\psi$. However, in our microscopic 
model, we have not found any situation in which the effective
parameters in eq 6 lead to a first-order transition.
The requirement of thermodynamic equilibrium with
the bulk imply strong limitations in the possible values
of these parameters.

\begin{figure}
\begin{center}
\epsfig{file=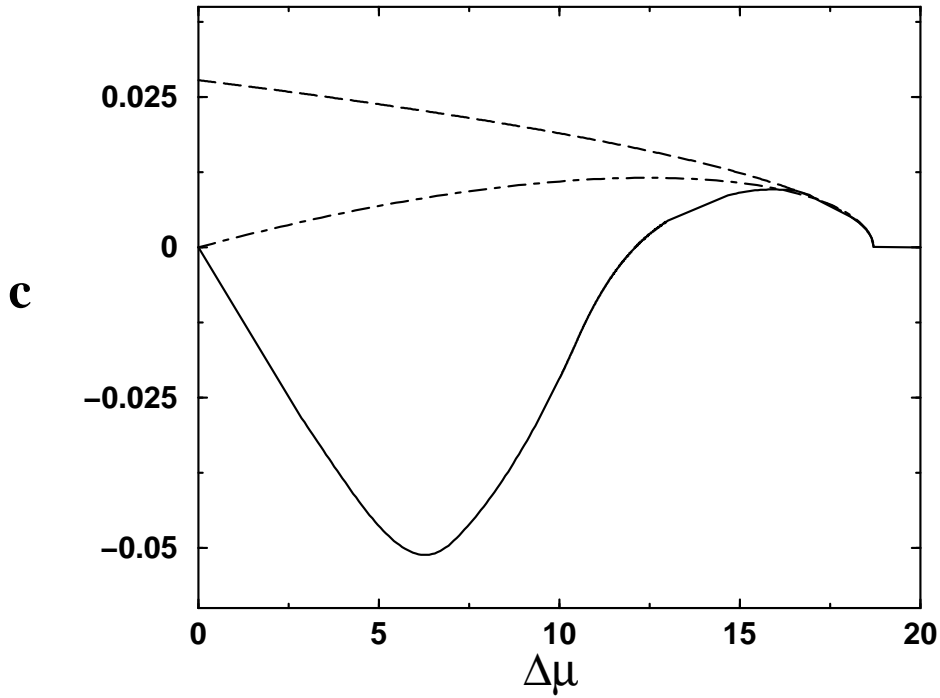,width=12cm}
\caption[]{
Curvature of the vesicle ${\sc c}$ as a function the difference between
chemical potentials $\Delta\mu$.
Solid line: our microscopic theory along of the equilibrium curve, with
$\tilde{K}=10$ and $\epsilon=1.4$.
Dashed line: schematic curve obtained with a Landau theory assuming
a coupling $ {\sc c} \phi $ between the curvature and the membrane composition.
Dot--dashed line: As for the previous curve, but now assuming a coupling
$ {\sc c} \phi \psi$.
}
\end{center}
\end{figure}

If we compare the empirical form of the
free energy given by the eq 6 with the numerical result for our microscopic 
model, we find that the 
fit of the free energy, as a function of curvature, to a second-order
polynomial is always very accurate. Thus confirming a
Helfrich--like
picture in which an expansion of the free energy to two second-order in
the curvature is carried out. The fit also allows us to
safely extrapolate to the infinite radius (zero curvature) limit.
However, the dependence of the 
free energy on the composition of the membranes,
i.e. with the variables $\psi$ and $\phi$, is quite beyond the
truncated Taylor expansion used in eq 6. The actual
results for the curvature, also shown in Figure 3
---solid line---,
show a richer dependence 
of the spontaneous curvature with $\Delta \mu$.

To obtain the global phase diagram for curved structures
in our model, we have taken as a starting point the results for the
planar membrane ${\sc c}=0$, presented in Figure 2.
In this way, for $\epsilon=1.4$, we
choose a value of the temperature, for instance $\tilde{K}=10$, 
so that for any value of 
$\Delta \mu$ there is a unique value of the
total bulk density of the surfactants, $\rho_o$, such that
the planar membrane has zero surface tension,
$\Delta\Omega(\rho_0,{\sc c}=0)=0$. When the dependence of
$\Omega$ with respect to the vesicle curvature, ${\sc c}$, 
is included we obtain results
like those in Figure 4 for 
$\Delta \mu=18.718$ (curve (a)) and $17.692$ (curve (b)).
In the first case,
the global minimum is precisely at ${\sc c}=0$ and the
membrane does not have spontaneous curvature. 
In case (b) the spontaneous curvature lowers
the value of $\Delta\Omega$ below zero. The global thermodynamic
equilibrium in this latter case requires a change of the
bulk density $\rho_0$, from $1.1761\ 10^{-6}$ to $1.1756\ 10^{-6}$
to recover the condition $\min[\Delta\Omega(\rho,{\sc c})]=0$,
as shown in curve (c). Now, repeating the procedure for any
values of $\tilde{K}$, we could draw 
the phase diagram for curved structures (vesicles),
but it is not necessary because is equal to the 
obtained for planar membranes since the shift of the bulk coexistence
density is too small to 
be distinguishable in the scale of Figure 2.

\begin{figure}
\begin{center}
\epsfig{file=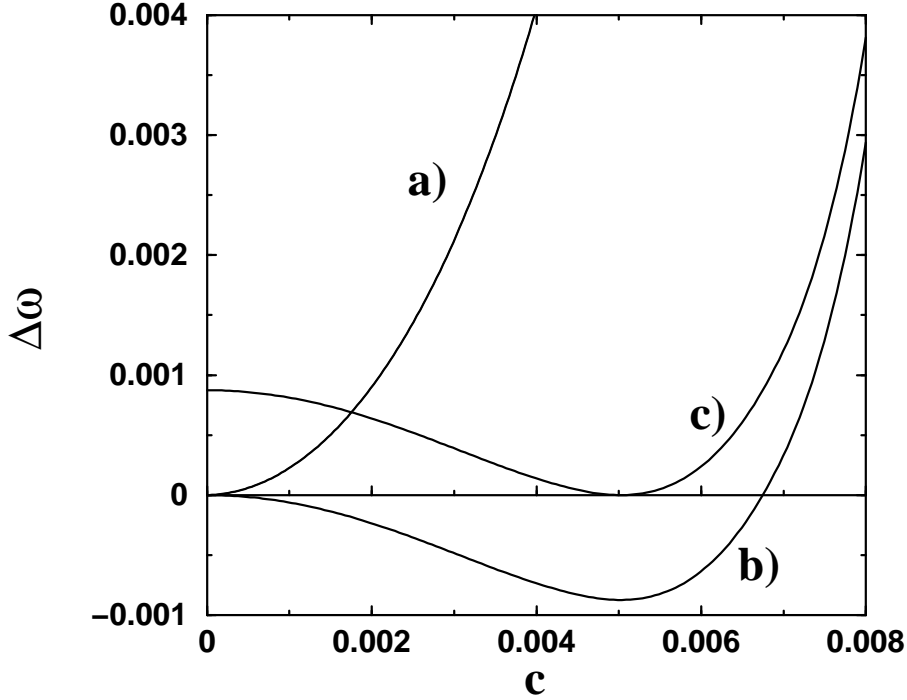,width=12cm}
\caption[]{
Excess free energy per particle,
$\Delta\omega=\Delta\Omega/\Delta N$, versus curvature for
$\tilde{K}=10$ and $\epsilon=1.4$.
a) Planar membrane $ {\sc c} = 0$ at coexistence, with $\log(\rho_0)=-13.5837$
and $\Delta\mu=18.718$;
b) Vesicle ${\sc c} \neq 0$ beyond coexistence, with $\log(\rho_0)=-13.6533$
and $\Delta\mu=17.692$;
c) Vesicle ${\sc c} \neq 0$ at coexistence, with $\log(\rho_0)=-13.6537$
and $\Delta\mu=17.692$.
Notice how slight the correction is between the parameters in b) and c).
}
\end{center}
\end{figure}

In the Figure 5 we present another view of the 
coexistence line for
bilayer vesicles, in terms of the concentration ratio and
of the total bulk density (always for 
$\tilde{K}=10$ and $\epsilon=1.4$). 
The full line gives the concentration
ratio measured in the bulk $log(\rho^{A}_{0}/\rho^{B}_{0})$
(which is precisely $\beta \Delta \mu$), moving along a horizontal
line would correspond to change the total concentration of
surfactant but keeping the bulk ratio between the two species.
Bulk dilutions on the left of the full line would be stable,
while on the right of the line they would form bilayer membranes.
This coexistence line also can be represented in term of the
composition of the membrane instead of the bulk composition. So,
in the same figure, 
the dashed lines give the ratio between the two surfactant species
in the membranes, within our parametrization
$ \rho^{\xi}_{m}= \eta_{1} x^{\xi}_{1} + \eta_{2} x^{\xi}_{2} $.
The ratio between the two species is directly given by the parameter $\psi$, 
$\rho^{A}_{m}/\rho^{B}_{m}= (1-2\psi)/(1+2\psi)$.
As already noted for the planar case,  
the condensation of the surfactant on the membrane 
produces a very important change in the relative
concentrations of the two surfactant species, given by the
vertical separation between the full and the dashed lines in the 
Figure 5. A bulk dilution in which the species $B$
is in clear minority (e.g. $\Delta \mu=\log(\rho_0^A/\rho_0^B) 
\sim 10$)
coexists with vesicles in which the two components have
comparable concentrations 
($\log(\rho_m^{A}/\rho_m^{B}) \sim 1$).

\begin{figure}
\begin{center}
\epsfig{file=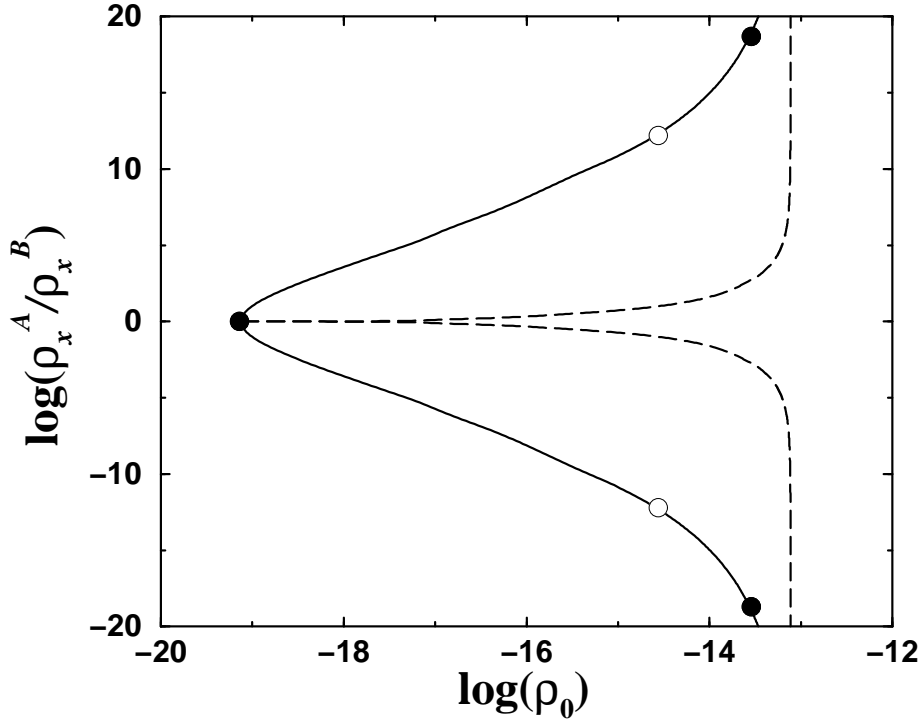,width=12cm}
\caption[]{
Equilibrium curve for the vesicles as a function of the
relative concentration and the total bulk density $\log(\rho_0)$. For the solid
line, the relative concentration is measured in the bulk
$\log(\rho_0^A/\rho_0^B) = \Delta\mu $. Alternately for the dashed line, the
relative concentration is measured in the membrane
$\rho^A_m/\rho^B_m=(1/2-\psi)/(1/2+\psi)$.
The energy parameters are
$\tilde{K}=10$ and $\epsilon=1.4$.
Notice how slowly the membrane ratio grows as compared with
the bulk one, reflecting the ability of
vesicles to concentrate the minority component.
As explained in the text,
the circles mark the places where ${\sc c}$ reaches zero.
}
\end{center}
\end{figure}

The results for the spontaneous curvature of the vesicles
along the coexistence line are given by the full line
in Figure 3, to be compared with the results of the
simple Landau theories.
For $ | \Delta \mu | > 18.7$ the
membrane is symmetric (and planar), since the
minority component is too scarce to allow the
segregation phase transition. This situation corresponds to the
behavior on the right hand side of Figure 5, 
in which both the bulk and the membrane tend asymptotically to the 
one-component case. As we lower the values of $ | \Delta \mu | $
the transition takes place, there appears spontaneous curvature
of the vesicles, and the composition asymmetry of the
two layers develops rapidly. But this transition requires the 
incorporation into
the membrane of a significant amount of the minority component, which
pushes the membranes towards more symmetric composition,
$\rho^A_m \approx \rho^B_m$, and it is represented by the 
dashed lines going to zero, on the left of Figure 5.
In accordance with the symmetry arguments presented above, the
membranes become flat again at the symmetric mixture
($\Delta \mu=0$ and hence $\psi=0$). The three points marked by
black circles along the coexistence line, in Figure 5,
correspond to these transitions from planar to spherical vesicles.
The existence of these transitions and the general shape 
of the phase diagram in Figure 5 is generic for 
symmetric surfactant mixtures and can be obtained from the 
simple Landau free energy (eq 6).
What is more amazing in our microscopic calculation
is the existence of intermediate points, at   
$\Delta \mu \approx \pm 12$, where the curvature 
also vanishes, although both $\phi$ and $\psi$ are nonzero.
These points, marked by open circles in Figure 5,
have unbalanced total composition both at the 
bulk and the membranes 
($\rho^A_0/\rho^B_0 \approx 10^{5}$, $\rho^A_m/\rho^B_m \approx 1.5$)
and the two layers of the membrane have broken the symmetry between 
the two species, but there still is no spontaneous curvature. 
This behavior is not predicted by simple Landau theories
and implies that the coupling between $\phi$ and ${\sc c}$ has
a complex dependence on $\psi$. 

In order to allow better visualization of the composition 
of the bilayer vesicle, we show
in Figure 6 the relative compositions of the A and B species
for each of the layers. 
For high values of $\Delta \mu > 18.7$, in the flat membrane regime,
we see the symmetry between the two layers, $x_1^\xi=x_2^\xi$.
As we decrease $\Delta \mu$ below the transition to spherical vesicles,
the curvature grows and  
the minority component B goes preferentially to the outer
layer of the membrane (labeled as 1), but even in that layer
the species B is still in minority. The curvature reaches a maximum 
for $\Delta \mu \approx 16$, with vesicles of radius about one
hundred molecular diameters.  For still lower
values of $\Delta \mu$ the vesicles become larger until for
$\Delta \mu \approx 12$ the spontaneous curvature disappears
again. At this point the composition of the outer layer in the 
vesicles has become nearly  symmetric,
$x_1^A \approx x_1^B$, while the inner layer keeps a clear majority of the
species A. When we go to further 
lower values of $\Delta \mu$ the membrane bends over in the 
opposite way, leaving now the layer with majority of component B
as the inner layer (indicated by negative curvatures in Figure
3 ). At $\Delta \mu \approx 6.3$ the size of the spontaneous
vesicles reaches the minimum, with a radius of about twenty molecular
diameters. The curvature goes again to zero if we 
approach the symmetric mixture $\Delta \mu=0$, while the composition
of the two layers is nearly frozen, with nearly complete segregation
of each species to different layers.

\begin{figure}
\begin{center}
\epsfig{file=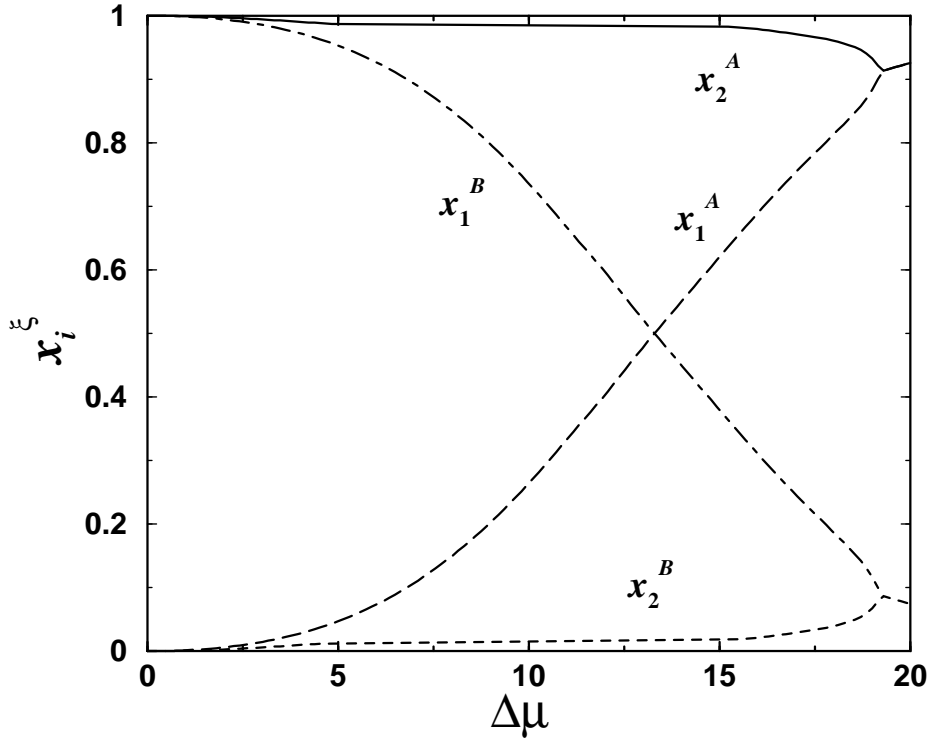,width=12cm}
\caption[]{
The relative compositions $x_i^\xi$ of each species
in each layer
versus $\Delta\mu$, along the coexistence line, for
$\tilde{K}=10$ and $\epsilon=1.4$.
}
\end{center}
\end{figure}

We have not found a direct 
explanation of this behavior, which may depend on the 
microscopic interactions of the model, but the general 
feature seems to be quite robust. The results for  
$\epsilon =1.2$ (i.e. decreasing the difference between the
$A$--$A$, $B$--$B$ and the $A$--$B$ interactions) show that
the range of $\Delta \mu$ with spontaneous curvature decreases
but there is still a change of sign
in ${\sc c}$, with the same qualitative trend 
as for the case with $\epsilon =1.4$.

Making use of the information presented in Figure 5 it is
possible to construct a phase diagram in terms of the usual
experimental variables, the total densities of each amphiphile 
in the dissolution:
$\rho_{t}^{\xi}=\rho_{o}^\xi+\rho_{m}^{\xi} \ a_m$, where
$a_m$ is the total amount of membrane area by unit volume of
the dissolution.
The two densities $\rho_{t}^{A}$ and $\rho_{t}^{B}$
determine a point in the diagram; if this point is located
on the left of the coexistence (solid) line, the bulk
densities are equal to the total ones; that is, no
structures are formed and $a_m=0$. If, on the other hand, the
point lies on the right, the molecules can either
stay on the bulk or form aggregates; 
the solution shifts towards a point on the coexistence line.
The final equilibrium point can be easily determined from the knowledge
of the composition of the aggregates, i.e. the dashed
line. With this information we redraw the phase diagram, Figure 7, in
the plane $\log(\rho_{t})$ -- $\log(\rho_{t}^A/\rho_{t}^B)$.
We can distinguish three zones: one in which no
aggregates are formed (I), another one in which they
form planar membranes(II), and a third one where they
form vesicles. The latter is divided in two subregions,
corresponding to positive (IIIa) and negative (IIIb)
curvature.
We believe these features of our model to be in qualitative agreement
with experiments with catanionic systems in the
dilute region \cite{DIA1,Kaler,DIA2,DIA3,DIA4},
specially the existence
of two regions where vesicles are stable (zones III) when the
mixture is not equivalent, separated by
a region in which the two components are present in similar amounts with
flat membranes forming and precipitating.

\begin{figure}
\begin{center}
\epsfig{file=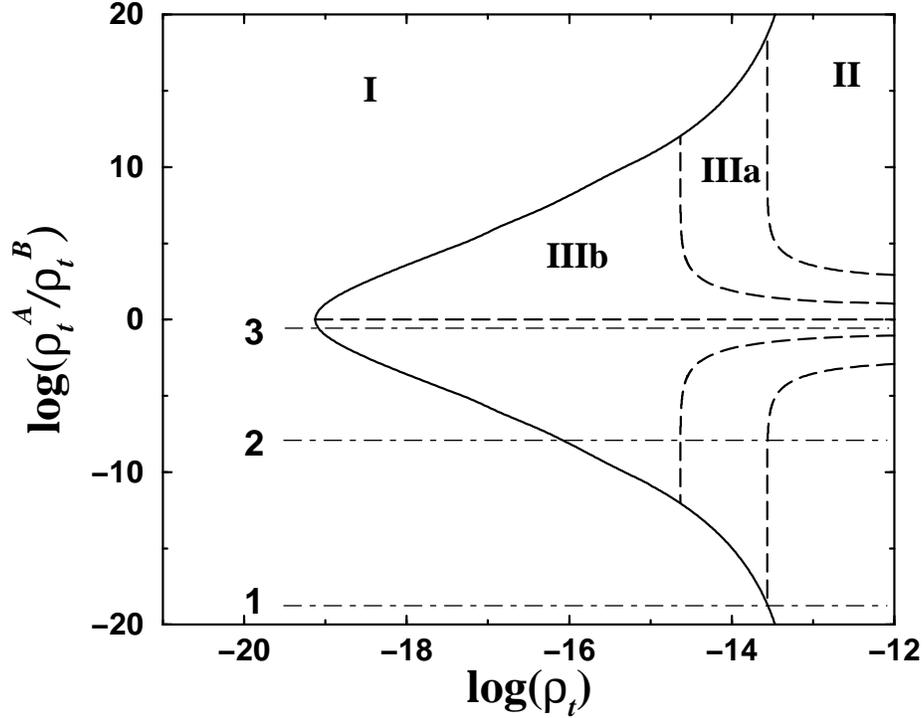,width=12cm}
\caption[]{
Phase diagram in the $\log(\rho_t)$ --  $\log(\rho_t^A/\rho_t^B)$
(which is not $\Delta\mu$, since now we refer to total
densities) plane,
at fixed $\tilde{K}=10$ and $\epsilon=1.4$.
The solid line is the coexistence line between free
molecules and aggregates; the dashed ones, between different types
of aggregates.
The meaning of the regions and the dot--dashed paths is
explained in the text.
The diagram is completely up--down symmetric, but for the sake of clarity
we only draw the regions in the upper side and the different paths
in the lower one.
}
\end{center}
\end{figure}

In order to compare our results with experiments which measure the evolution of 
the average radius of the vesicle when water is added to 
the solution \cite{Kaler},
we have consider the behavior of our system when the ratio of the
two components, $\rho_t^A/\rho_t^B$,
is kept fixed while the total density
$\rho_t=\rho_t^A+\rho_t^B$ is varied. This
process would follow horizontal lines in Figure 
7. Depending of the value of $\log(\rho_{t}^A/\rho_{t}^B)$ 
we can have different behavior; we show three interesting
examples, labeled from 1 to 3. The sequence of structures, from low
to high values of $\rho_t$, would be: for the first case, 
no structures--planar membranes; for the second, 
no structures--vesicles--planar membranes; 
and for the third, no structures--vesicles. 

\begin{figure}
\begin{center}
\epsfig{file=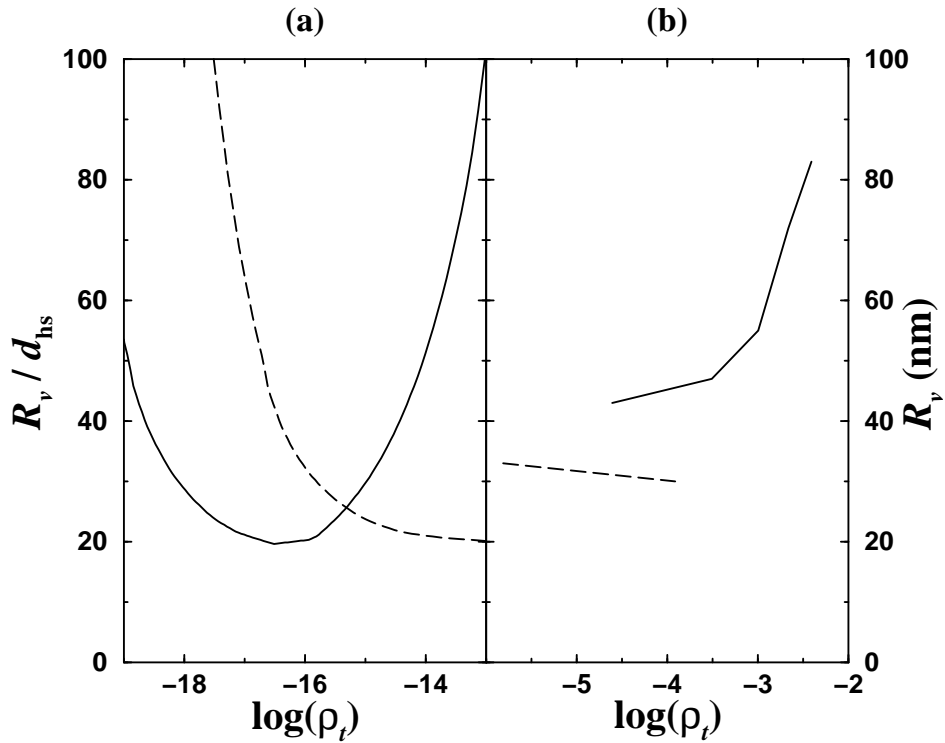,width=12cm}
\caption[]{
Average vesicle radius as a function of the total
density, keeping the composition ratio of the two components fixed.
In panel (a) we present the results obtained with
our model, following two horizontal paths
of type 3 in Figure 7,
as a function of $\log(\rho_{t})$. The full line corresponds to
$\log(\rho_{t}^A/\rho_{t}^B) = -1 $, and  the
dashed line to $ -0.1 $.
Panel (b) show the experimental
results of Kaler {\sl et al} in ref 2, for composition ratios $1/9$
(solid line) and
$1/4$ (dashed line). The radii are given in
nanometers and the total density is measured
by $\log(1-w/100)$, where $w$ is the water-weight percentage.
}
\end{center}
\end{figure}

The evolution of the radius with 
$\rho_t$ depends strongly on the
ratio $\rho_t^A/\rho_t^B$ and it may be nonmonotonic.
In Figure 8(a) we present our results for two different
cases, both corresponding to a sequence of
structure type 3. The nonmonotonic dependence comes directly from the
nonmonotonic behavior of ${\sc c}$ in Figure 3; it would
not appear in a Landau free energy analysis with a constant
coupling between ${\sc c}$ and $\phi$. The experimental observation
of the spontaneous curvature presents strong technical difficulties,
with uncertainties in the equilibration and the polydispersity.
However, the results, for mixtures of
cetyl trimethilammonium tosylate and sodium
dodecylbenzene sulfonate \cite{Kaler}, shown in the panel (b) of the Figure
8, present clear examples of non-monotonic dependence of
the radius with the total concentration of surfactant, $\rho_t$.
The fragmentary experimental information does not allow us to 
confirm or to refute the presence of the intermediate point
with ${\sc c}=0$ found in our model. A direct comparison with our 
predictions would require the chemical analysis of the inner and 
the outer layers of the vesicles, which has not been carried out so far.
As for the typical size of the spontaneous vesicles, we can only expect a
qualitative agreement with our simplified model. In our results,
the vesicles have always radius larger than $20$ and typically
in the range of $100$ in units of the molecular size (the hard 
core parameter); the experimental values ranging between $20$ and $80 \ nm$
are in the same order of magnitude, for molecular sizes in the 
nanometer range.

\section{Conclusions} 

We have developed a self-consistent 
microscopic model for the study of aggregates formed in a
mixture of two different amphiphilic species ($A$ and $B$) in water. 
The simplicity of the
interactions considered here makes possible
a self-consistent thermodynamic treatment,
assuring thermodynamic equilibrium between the solution
of amphiphiles and the aggregates. We have assumed that the interactions
between the amphiphiles are symmetric, the $A$--$A$ interaction is equal to
the $B$--$B$ one but different from the $A$--$B$ interaction.
Using a semiempirical Landau theory, we ---
as Safran {\it et al} did previously \cite{Safran1,Safran2}--- 
have found
that the mixture of interacting amphiphiles stabilizes the
spherical vesicle with respect to planar membranes. However, there are
important differences between the two models: Safran {\it et al} use a Landau
free energy based on
the hypothesis that the driving force for the spontaneous
curvature is the existence of independent phase transitions
in each monolayer, due to an effective intralayer repulsion
between the molecules of different types.  In our case,
we assume that the driving force is an interlayer attraction
between different molecules in opposite layers of the membrane.
Our hypothesis seems to be easier to justify for real
catanionic mixtures and leads to a different coupling
between the asymmetry of the molecular composition and the
curvature of the vesicles. In our case, the coupling term go like
$ {\sc c} \phi f(\psi) $, where f is an odd function,
thus it vanishes
not only for membranes with equal composition in the two layers
($\phi=0$) but also for membranes with different layers $\phi \neq 0$
but with equal global compositions of the two surfactants ($\psi=0$).

Moreover, in our density functional approach, we obtain
the free energy directly from molecular interactions, instead
of having empirical parameters as in any Landau free energy.
Our results show that the presence of first-order phase
transitions, predicted by the different Landau theories,
may be easily frustrated by the requirements of global
thermodynamic stability. The membranes may be regarded as
two dimensional phases, but they have to be at coexistence
with a diluted bulk and, in our microscopic model, this
imposes severe restrictions in the possible behavior of
the free energy. In fact, the truncated expansion of the
free energy in terms of the order parameters $\phi$ and $\psi$
turns out not no be so useful, since the coefficients vary in
a quite complicated manner as the thermodynamic conditions are
varied --- despite the truncation being, of course, correct in 
a limited range
of application, i.e. small values of the order parameters.

The phase diagram obtained with our model is in qualitative agreement
with the experimental diagram for a system comprised of an anionic 
and a cationic surfactant in the dilute region. 
We have also studied the
evolution of the vesicles size when water is added to the dissolution.
We have found that the radius of the vesicles increase or decrease 
depending of the ratio $\rho^{A}/\rho^{B}$. This fact could
explain the complicated experimental evolution of the vesicle radius,
although experimental technical difficulties can also have an
influence.

In conclusion, the minimal model for the mixed anionic-cationic
surfactant vesicles, presented in this work, is a fundamental 
description of the thermodynamic of these systems. Although
the micellar phase has not been taken into account in the present work, 
this fact does not change the main conclusions since
the vesicles and plane membranes are always more stable
than the micelles. This assumption is based on the fact that
we have used  
$q=0.5$ for all the intermolecular potentials, and this value
stabilizes the plane membranes compared to   
the micelles for the monocomponent case.  Also, we have
neglected the entropy of shape fluctuations and the 
center of mass translations for the aggregates. The inclusion of 
such terms for vesicles would give a finite width to the
'coexistence line' calculated here. However, given the large
size of the spontaneous vesicles (always with more than $10^4$
molecules) and the large value of the bending rigidity, the effects
would be negligible. Extension of the
present model to the study of mixtures of amphiphiles with
different self-assembly behavior, including micellar aggregates,
is being in progress.

\section*{Acknowledgments}
The continuous interest of Prof. A. Somoza in this work and useful 
suggestions by Prof. John P. Hernandez are gratefully acknowledged.
This work has been supported by the Direcci\'on General de Investigaci\'on
Cient\'{\i}fica y T\'ecnica of Spain, under grant number PB94-005-C02
and the Comunidad Aut\'onoma de Madrid, under grant F.P.I.--1995.


\begin{thebibliography}{99}
\bibitem{Safran2} 
MacKintosh, F.C.; Safran, S.A.
{\sl Phys. Rev. E} {\bf 1993}, {\it 47}, 1180. Note that the definition of $\psi$
here differs from the previous reference; ours agrees with this article.
\end{thebibliography}
\end{document}